\documentclass[prd,aps,nofootinbib,onecolumn,showpacs,12pt]{revtex4-1}
\usepackage{graphicx,epsfig}

\def\beq{\begin{equation}}
\def\eeq{\end{equation}}
\def\bea{\begin{eqnarray}}
\def\eea{\end{eqnarray}}
\def\beeq{\begin{eqnarray}}
\def\eeeq{\end{eqnarray}}

\def\nnb{\nonumber}

\def\rar{\rightarrow}
\def\nnb{\nonumber}

\def\ba{\begin{array}}
\def\ea{\end{array}}

\def\xis0{{\Xi^{*0}}}

\def\g5{\gamma_5}

\def\ar{&+& \!\!\!}

\usepackage{epsf}
\usepackage{graphicx}

\def\ar{&+& \!\!\!}

\def\nnb{\nonumber}

\begin{document}

\title{ \bf  B to strange tensor meson transition in a model with one universal extra dimension}
\author{  N. Kat{\i}rc{\i}$^{\dag}$, K. Azizi$^{\ddag}$\\
 Physics Division,  Faculty of Arts and Sciences,
Do\u gu\c s University,
 Ac{\i}badem-Kad{\i}k\"oy, \\ 34722 Istanbul, Turkey\\
$^\dag$e-mail:nkatirci@dogus.edu.tr\\
$^\ddag$e-mail:kazizi@dogus.edu.tr}

\begin{abstract}
We analyze the semileptonic $B\to K_2^*(1430)l^+l^-$  transition  in universal extra dimension model. In particular, we present
 the sensitivity of related observables such as branching ratio, polarization distribution and forward-backward asymmetry to the
compactification factor ($1/R$) of extra dimension. The obtained results from extra dimension model show overall a considerable deviation from the standard model  predictions
 for small values
 of the compactification factor. This can be considered as an indication for existence of extra dimensions.
\end{abstract}
\pacs{ 12.60-i, 13.20.-v , 13.20.He}

\maketitle

\section{Introduction}
The semileptonic  $B$ meson decays are important frameworks to restrict the Standard Model (SM) parameters as well as search for new physics beyond the SM. Experimental progress at $B$ 
factories offers the possibilities to study such decay channels in near future (see for instance \cite{belle,babar,LHC,superb,lhcb}).  Among the semileptonic $B$ decays, the 
 $B\to K_2^*(1430)l^+l^-$ transition is important as it happens via loop  flavor changing neutral current (FCNC) of $b \to s$ transition at quark level. Such loop transition can be used to 
explore the effects originating from new physics beyond the SM, hence, theoretical calculations of the related parameters to these transitions become important in this respect.

Universal extra dimension (UED) model is one of the popular
extension of the SM.  This model is a  category of the  extra dimension (ED) \cite{antoniadis1,antoniadis2,arkani} which allows
the SM fields (both gauge bosons and fermions) to propagate in the extra dimensions. Comparison of the results obtained by UED with those of the SM can offer interesting phenomenology.
 We consider the
simplest UED model where just a single universal extra dimension
compactified on a circle of radius $R$  called the
Appelquist, Cheng and Dobrescu (ACD) model \cite{acdd} to investigate the  $B\to K_2^*(1430)l^+l^-$  transition.
The effective Hamiltonian responsible for  $b \to s$ transition was calculated in
the ACD model in \cite{Buras:2002ej,R7623,R7626,R7627,R762777}. In this model, the Kaluza-Klein (KK)
particles interact with themselves as well as with the SM particles. These interactions bring
 additional Feynman diagrams compared with the SM and require changes in the  Wilson coefficients entering to the effective Hamiltonian. In this model, the Wilson coefficients and as a result, the effective Hamiltonian 
 are described in terms of the compactification factor $1/R$.

The main ingredients in analysis of the considered transition both in UED and SM models are form factors entered to the   transition matrix elements.  These form factors  have been recently calculated
 both in the perturbative QCD  \cite{wang} and  light cone QCD sum rules \cite{zhigang} approaches. Using the corresponding form factors, we depict sensitivity of the related physical observables 
such as branching ratio, polarization distribution and
forward-backward asymmetry to the
compactification factor $1/R$ and compare the obtained results from extra dimension with  those  of the SM.  The ACD model has been previously applied to investigate the following channels:
$\Lambda_b\rightarrow\Lambda \nu\bar\nu$ and
$\Lambda_b\rightarrow\Lambda \gamma$ \cite{R7624,wangying},
$\Lambda_b\rightarrow\Lambda l^+l^-$ \cite{R7601,wangying}, $B\to K^*l^+l^-$,
$B\to K^*\nu^+\nu^-$, and $B\to K^*\gamma$ \cite{fazio} and  $B\to
K_0^*(1430)l^+l^-$ \cite{sirvanli}. Recently, we have also investigated many observables describing the $\Lambda_b\rightarrow\Lambda l^+l^-$ 
transition using the corresponding form factors obtained from full QCD in UED model \cite{kank1}. For some other applications of the ACD model to hadron physics see \cite{yee,bashiry,carlucci,aliev,ahmet}. Note that the 
 $B\to K_2^*(1430)l^+l^-$ transition has also been investigated in the standard model and two
new physics scenarios: vector-like quark model and family non-universal $Z^ \prime$ model \cite{li}.

The outline of the paper is as follows. In next section, we introduce the effective Hamiltonian responsible for the considered transition. In section III, the transition matrix elements and fit functions of the 
responsible form factors are presented. In section IV, we discuss the sensitivity of the aforementioned physical quantities to the compactification factor $1/R$ and compare
 the obtained results with the SM predictions. Last section is devoted to our conclusions.

\section{Effective Hamiltonian responsible for  the $B\to K_2^*(1430)l^+l^-$  transition }
 The $B\to K_2^*(1430)l^+l^-$ transition proceeds via  loop  $b \to s$ transition whose effective Hamiltonian  is written as:
\bea \label{e8401} {\cal H}^{eff} &=& {G_F \alpha_{em} V_{tb}
V_{ts}^\ast \over 2\sqrt{2} \pi} \Bigg[ C_9^{eff}
\bar{s}\gamma_\mu (1-\gamma_5) b \, \bar{\ell} \gamma^\mu
\ell + C_{10} \bar{s} \gamma_\mu (1-\gamma_5) b \,
\bar{\ell} \gamma^\mu
\gamma_5 \ell \nnb \\
&-&  2 m_b C_7^{eff}  {1\over q^2}\bar{s} i
\sigma_{\mu\nu}q^{\nu} (1+\gamma_5) b \, \bar{\ell} \gamma^\mu
\ell \Bigg]~. \eea
 where $G_F$ is the Fermi coupling constant,  $V_{ij}$ are elements of
the Cabibbo-Kobayashi-Maskawa (CKM) matrix,  $\alpha_{em}$  is the fine structure
constant and $C_7^{eff}$, $C_9^{eff}$ and $C_{10}$  are 
Wilson coefficients. 
 The  Wilson coefficients in ACD Model  are calculated in
\cite{Buras:2002ej,R7623,R7626,R7627,R762777} in
leading logarithmic approximation. In this model,
each Wilson coefficient is described in terms of  some periodic
functions $F(x_{t},1/R)$  having an ordinary SM part  $F_0 (x_t )$ and an extra part in terms of the
compactification factor $1/R$, i.e., 
\bea F(x_t,1/R)=F_0(x_t)+\sum_{n=1}^{\infty}F_n(x_t,x_n).
\label{functions} \eea 
Here  $x_{t}=m_{t}^{2}/M_{W}^{2}$ and $m_t$ is
the top quark mass.
In the above equation, $x_n=m_n^2/m_W^2$ with  $m_n=n/ R$ being  the mass of the KK particles and $n=0$ corresponds to the ordinary SM  particles. 
 The Glashow-Illiopoulos-Maiani (GIM) mechanism guaranties the finiteness of the functions $F(x_t,1/R)$ and satisfies the 
condition $F(x_t,1/R)\rar F_0(x_t)$ when $R\rar 0$. Explicit expressions for the Wilson coefficients with all input parameters are presented  in \cite{kank1}. From the expressions for the Wilson coefficients, we see
that, the $C_7^{eff}$ and  $C_{10}$ are only functions of the compactification factor. However, the $C_9^{eff}$, besides the $1/R$, depends also on the transferred momentum squared $q^2$. 
Using the explicit expressions for these coefficients, we obtain the numerical
 values for $C_7^{eff}(1/R)$, $C_{10}(1/R)$ as well
as  $C_9^{eff}(1/R,q^2)$ at a fixed value of $q^2$ and different values of $1/R$ in
Table~\ref{SCXY}. In this Table, we also present the values of these coefficients from the SM.  Here, we would like to make the following comment about the range of the compactification factor, $1/R$. 
 From the electroweak precision tests, the lower limit for $1/R$ is obtained as $%
250~GeV$  if $M_h\geq250~GeV$ expressing larger KK contributions to the low energy FCNC processes,
 and $300~GeV$  if $M_h\leq250~GeV$ \cite{acdd,yee}. In the present study, we  consider the range of  $1/R$ from $200~GeV$ up
to $1000~GeV$. With a quick glance at  Table~\ref{SCXY}, we observe that
 the values of Wilson coefficients in UED model  differ considerably from their
SM values.  In particular,  $C_{10}$ is enhanced and 
$C_7^{eff}$ is suppressed.
\begin{table}[hbt]
\begin{center}
\begin{tabular}{|c||c|c|c|}\hline
  $1/R~[{\rm GeV}]$   & $C_7^{eff}$ & $C_{10}$ & $C_9^{eff}(14)$
 \\\hline
$ 200$  &
 $-0.195212$ & $-5.61658$ &
 $4.83239 + 3.59874i$\\\hline
$ 300$ &
 $-0.244932$ & $-4.92684$ &
 $4.77624 + 3.55939i$ \\\hline
$400$  & $ -0.266419$ & $-4.65118$ & $4.7538 + 3.54366i$\\\hline
$500$ & $-0.277351$ & $ -4.51581$ & $ 4.74278 + 3.53594i$\\\hline
 $600$  & $-0.283593$ & $-4.43995$ & $ 4.7366 + 3.53161i$\\\hline
 $700$   & $-0.287468$ &
$-4.39337$ & $4.73281 + 3.52895i$\\\hline $800$  & $-0.29003$ &
$-4.36279$ & $4.73032 + 3.52721i$\\\hline
 $900$   & $ -0.291808$ &
$-4.34166$ & $4.7286 + 3.526i$\\\hline
 $1000$  & $-0.293092$ & $-4.32646$ & $4.72736 + 3.52514i $\\\hline
 SM      &
$-0.298672$ & $-4.26087 $ &
 $4.72202 + 3.52139i$ \\\hline
\end{tabular}
\end{center}
\caption[]{\small Numerical values for $C_7^{eff}$,   $C_{10}$ and
values of $C_9^{eff}$ at $q^2=14$   for different $1/R$'s and the
SM. \label{SCXY}}
\end{table}

\section{Transition matrix elements and $B$ to tensor meson form factors}
To obtain the physical quantities, we need to know the amplitudes defining the considered transition.  The decay  amplitude for $B\to Tl^+l^-$ are obtained sandwiching the  effective
Hamiltonian between the initial and final states:
\begin{eqnarray} \label{amp} 
 \langle T(P_2,\epsilon)|{\cal H}^{eff}|B(P_B)\rangle
\end{eqnarray}
Where, $P_2$ and $P_B$ are the momenta of the final and initial states, respectively, $\epsilon_\mu=\frac{1}{m_B}\epsilon_{\mu\nu} P_B^\nu$ and $\epsilon_{\mu\nu}$ is polarization tensor of
 the   tensor meson.
To proceed, we need to know the following matrix elements which are parameterized in terms of form factors~\cite{Hatanaka:2009gb,Hatanaka:2009sj,wang,Yang:2010qd}:
 \begin{eqnarray}\label{defmat}
  \langle T(P_2,\epsilon)|\bar s\gamma^{\mu}b| B(P_B)\rangle
   &=&-\frac{2V(q^2)}{m_B+m_{T}}\epsilon^{\mu\nu\rho\sigma} \epsilon^*_{\nu}  P_{B\rho}P_{2\sigma}, \nonumber\\
  \langle T(P_2,\epsilon)|\bar s\gamma^{\mu}\gamma_5 b| B(P_B)\rangle
   &=&2im_{T} A_0(q^2)\frac{\epsilon^* \cdot  q }{ q^2}q^{\mu}
    +i(m_B+m_{T})A_1(q^2)\left[ \epsilon^*_{\mu}
    -\frac{\epsilon^* \cdot  q }{q^2}q^{\mu} \right] \nonumber\\
    &&-iA_2(q^2)\frac{\epsilon^* \cdot  q }{  m_B+m_{T} }
     \left[ P^{\mu}-\frac{m_B^2-m_{T}^2}{q^2}q^{\mu} \right],\nonumber\\
 \langle T(P_2,\epsilon)|\bar s\sigma^{\mu\nu}q_{\nu}b| B(P_B)\rangle
   &=&-2iT_1(q^2)\epsilon^{\mu\nu\rho\sigma} \epsilon^*_{\nu} P_{B\rho}P_{2\sigma},\nonumber\\
  \langle T(P_2,\epsilon)|\bar s\sigma^{\mu\nu}\gamma_5q_{\nu}b| B(P_B)\rangle
   &=&T_2(q^2)\Bigg[(m_B^2-m_{T}^2) \epsilon^*_{\mu}
       - {\epsilon^* \cdot  q }  P^{\mu} \Bigg] \nonumber\\
       &&+T_3(q^2) {\epsilon^* \cdot  q }\left[
       q^{\mu}-\frac{q^2}{m_B^2-m_{T}^2}P^{\mu}\right], 
      \label{eq:BtoTformfactors-definition}
 \end{eqnarray}
where $q=P_B-P_2$, $ P=P_B+P_2$, and $V,~A_{0,1,2}$ and $T_{1,2,3}$ are form factors. At point, $q^2=0$, we  have the relation,
$2m_{T}A_0(0)=(m_B+m_{T})A_1(0)-(m_B-m_{T})A_2(0)$ in order to cancel the pole at $q^2=0$.

The form factors of $B\to T$ transition are calculated in \cite{wang} using the perturbative QCD and we use them in our analysis.  The form factors are best extrapolated by \cite{wang}:
\begin{eqnarray}
 F(q^2)&=&\frac{F(0)}{(1-q^2/m_B^2)(1-a(q^2/m_B^2)+b(q^2/m_B^2)^2)},\label{eq:fit-B-T}
\end{eqnarray}
where, the  parameters $a,b$ and $F(0)$ for form factors $V,~A_{0,1}$ and $T_{1,2,3}$ are presented in
Table~\ref{Tab:formfactorsBtoTbeforemixing}. Neglecting higher power corrections, $A_2$ is related to
$A_0$ and $A_1$ by:
\begin{eqnarray}
A_2(q^2)=\frac{m_B+m_{T}}{m^2_B-q^2}\Bigg[(m_B+m_{T})A_1(q^2)-2m_{T}A_0(q^2)\Bigg]. 
\end{eqnarray}

\begin{table}
\caption{ Parameters entering to the fit function of the form factors responsible for $B\to T$ transition.}
 \label{Tab:formfactorsBtoTbeforemixing}
 \begin{center}
 \begin{tabular}{ c c ccc}
\hline
 $F$       & $F(0)$  & $a$ &$b$    \\ \hline
   $V^{B\rightarrow K_{2}^*}$    & $0.21_{-0.04-0.03}^{+0.04+0.05}$
                    & $1.73_{-0.02-0.03}^{+0.02+0.05}$
                    & $0.66_{-0.05-0.01}^{+0.04+0.07}$  \\ \hline
 $A_0^{B\rightarrow K_{2}^*}$  & $0.18_{-0.03-0.03}^{+0.04+0.04}$
                    & $1.70_{-0.02-0.07}^{+0.00+0.05}$
                    & $0.64_{-0.06-0.10}^{+0.00+0.04}$ \\ \hline
 $A_1^{B\rightarrow K_{2}^*}$  & $0.13_{-0.02-0.02}^{+0.03+0.03}$
                    & $0.78_{-0.01-0.04}^{+0.01+0.05}$
                    & $-0.11_{-0.03-0.02}^{+0.02+0.04}$  \\ \hline
 $T_1^{B\rightarrow K_{2}^*}$  & $0.17_{-0.03-0.03}^{+0.04+0.04}$
                    & $1.73_{-0.03-0.07}^{+0.00+0.05}$
                    & $0.69_{-0.08-0.11}^{+0.00+0.05}$  \\ \hline
 $T_2^{B\rightarrow K_{2}^*}$  & $0.17_{-0.03-0.03}^{+0.03+0.04}$
                    & $0.79_{-0.04-0.09}^{+0.00+0.02}$
                    & $-0.06_{-0.10-0.16}^{+0.00+0.00}$    \\ \hline
 $T_3^{B\rightarrow K_{2}^*}$  & $0.14_{-0.03-0.02}^{+0.03+0.03}$
                    & $1.61_{-0.00-0.04}^{+0.01+0.09}$
                    & $0.52_{-0.01-0.01}^{+0.05+0.15}$   \\ \hline
 \end{tabular}
 \end{center}
 \end{table}


\section{Some Observables Relevant to the $B\to T l^+l^-$ Transition}
In this section, we present sensitivity of some physical quantities to the compactification factor and compare the obtained results from the UED with  SM predictions.
\subsection{Differential Decay Rate and Branching Ratio}
Using the amplitude from Eq. (\ref{amp})   and definitions of the transition matrix elements  in terms of the form factors from Eq. (\ref{defmat}), the differential decay rate  is obtained as \cite{li}:
\begin{eqnarray}
 \frac{ d\Gamma(q^2,1/R)}{dq^2}&=& \frac{1}{4}\Bigg[3I_1^c(q^2,1/R)+6I_1^s(q^2,1/R)-I_2^c(q^2,1/R)-2I_2^s(q^2,1/R)\Bigg],
\end{eqnarray}
 where,

 \bea I_1^c(q^2,1/R)&=&  \Bigg[|A_{L0}(q^2,1/R)|^2+|A_{R0}(q^2,1/R)|^2\Bigg]+8\frac{m_l^2}{q^2}
 {\rm Re}\Bigg[A_{L0}(q^2,1/R)A^*_{R0}(q^2,1/R) \Bigg]\nnb \\
\ar~ 4\frac{m_l^2}{q^2} |A_t(q^2,1/R)|^2, \nnb \\
I_1^s(q^2,1/R)&=&\frac{3}{4}
\Bigg[|A_{L\perp}(q^2,1/R)|^2+|A_{L||}(q^2,1/R)|^2  \nnb \\
\ar~ |A_{R\perp}(q^2,1/R)|^2
 +|A_{R||}(q^2,1/R)|^2\Bigg](1-\frac{4m_l^2}{3q^2}) \nnb \\
\ar~ \frac{4m_l^2}{q^2} {\rm
Re}\Bigg[A_{L\perp}(q^2,1/R)A_{R\perp}^*(q^2,1/R)
 + A_{L||}(q^2,1/R)A_{R||}^*(q^2,1/R)\Bigg], \nnb \\
I_2^c(q^2,1/R)  &=& -v^2\Bigg[  |A_{L0}(q^2,1/R)|^2+
 |A_{R0}(q^2,1/R)|^2\Bigg], \nnb \\
I_2^s(q^2,1/R) &=&
\frac{1}{4}v^2\Bigg[|A_{L\perp}(q^2,1/R)|^2+|A_{L||}(q^2,1/R)|^2 \nnb\\
\ar~ |A_{R\perp}(q^2,1/R)|^2+|A_{R||}(q^2,1/R)|^2\Bigg], \eea
 and
\bea 
v&&=\sqrt{1-4m_l^2/q^2},\nonumber\\
A_{L0}(q^2,1/R) &&= N_{K^*_2}(q^2)
\frac{\sqrt{\lambda}}{\sqrt6 m_Bm_{K_2^*}}\frac{1}{2m_{K^*_2}\sqrt
{q^2}} \Bigg[(C_9^{eff}(q^2,1/R)-C_{10}(1/R)) \nonumber\\ &&
[(m_B^2-m_{K^*_2}^2-q^2)(m_B+m_{K^*_2})A_1(q^2)-\frac{\lambda}{m_B+m_{K^*_2}}A_2(q^2)]\nonumber\\&&
 +2m_b(C_{7L}^{eff}(1/R)-C_{7R}^{eff}(1/R))[
(m_B^2+3m_{K_2^*}^2-q^2)T_2(q^2) \nonumber \\&&
-\frac{\lambda}{m_B^2-m_{K_2^*}^2}T_3(q^2)]\Bigg],\nonumber\\ 
 A_{L\perp}(q^2,1/R)&&= -\sqrt{2}
 \frac{\sqrt{\lambda}}{\sqrt8m_Bm_{K_2^*}}N_{K_2^*}(q^2)\Bigg[[C_9^{eff}(q^2,1/R)-C_{10}(1/R)]\frac{\sqrt \lambda V(q^2)}{m_B+m_{K^*_2}} \nonumber \\ &&
 +\frac{2m_b(C_{7L}^{eff}(1/R)+C_{7R}^{eff}(1/R))}{q^2}\sqrt \lambda T_1(q^2)\Bigg],\nonumber\\
  A_{L||}(q^2,1/R)&&= \sqrt{2}\frac{\sqrt{\lambda}}{\sqrt
  8m_Bm_{K_2^*}}N_{K_2^*}(q^2)\Bigg[[C_9^{eff}(q^2,1/R)-C_{10}(1/R)] (m_B+m_{K^*_2})A_1(q^2) \nonumber \\ &&
  +\frac{2m_b(C_{7L}^{eff}(1/R)-C_{7R}^{eff}(1/R))}{q^2}(m_B^2-m_{K^*_2}^2)T_2(q^2)\Bigg],\nonumber\\
A_t(q^2,1/R)&&= 2N_{K^*_2}(q^2) \frac{\sqrt{\lambda}}{\sqrt6m_Bm_{K_2^*}}C_{10}(1/R)\frac{\sqrt \lambda}{\sqrt
 {q^2}}A_0(q^2),\nonumber\\
N_{K_2^*}(q^2)&&=\Bigg[\frac{G_F^2 \alpha_{\rm em}^2}{3\cdot 2^{10}\pi^5
m_B^3}|V_{tb}V_{ts}^*|^2 q^2\lambda^{1/2}
v{\cal B}(K_2^*\to
K\pi)\Bigg]^{1/2}.
  \eea 
In the above equations, ${\cal B}(K_2^*\to K\pi)=0.499\pm0.012$ \cite{Amsler:2008zz} and
$\lambda=\lambda(m^2_{B},m^2_{K_2^*}, q^2)$ with $
\lambda(a^2,b^2,c^2)=(a^2-b^2-c^2)^2-4b^2c^2$.  
The right-handed  amplitudes are obtained via
\begin{eqnarray}
 A_{Ri}(q^2,1/R)
  &=& A_{Li}(q^2,1/R)|_{C_{10}(1/R)\to -C_{10}(1/R)},
\end{eqnarray}
where, $i=0,\perp$ or $||$.

Integrating  the differential decay rate over $q^2$ in the allowed
physical region, i.e.  $4 m_\ell^2\le q^2 \le
(m_{B}-m_{K^*_{2}})^2$, the $1/R$ dependent total decay
width is obtained. Using the lifetime   of the B meson, $\tau_{B}=1.530\times
10^{-12}~s$, and the input parameters, $m_b=4.8~GeV$, $|V_{tb}V_{ts}^\ast|=0.041$,
 $G_F = 1.17 \times 10^{-5}~
GeV^{-2}$, $\alpha_{em}=\frac{1}{137}$, 
   $m_{B}=5.28, 
m_{K_2^*}=1.43 {~ GeV}$,  $m_\mu = 0.1056~GeV$
and $m_\tau = 1.771~GeV$ \cite{Amsler:2008zz}, we acquire the $1/R$ dependent   branching
ratios as presented in Fig. 1. Note that we consider the uncertainties related to the hadronic form factors given in Table~\ref{Tab:formfactorsBtoTbeforemixing}  in our plots.
\begin{figure}[h!]
\label{fig1} \centering
\begin{tabular}{cc}
\epsfig{file=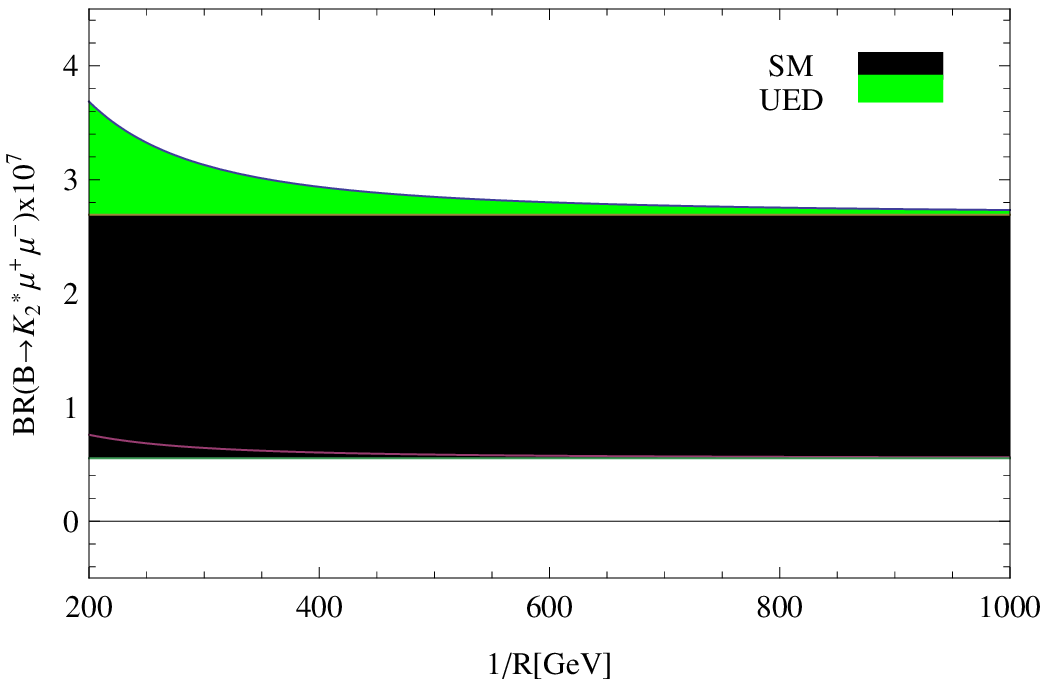,width=0.45\linewidth,clip=} &
\epsfig{file=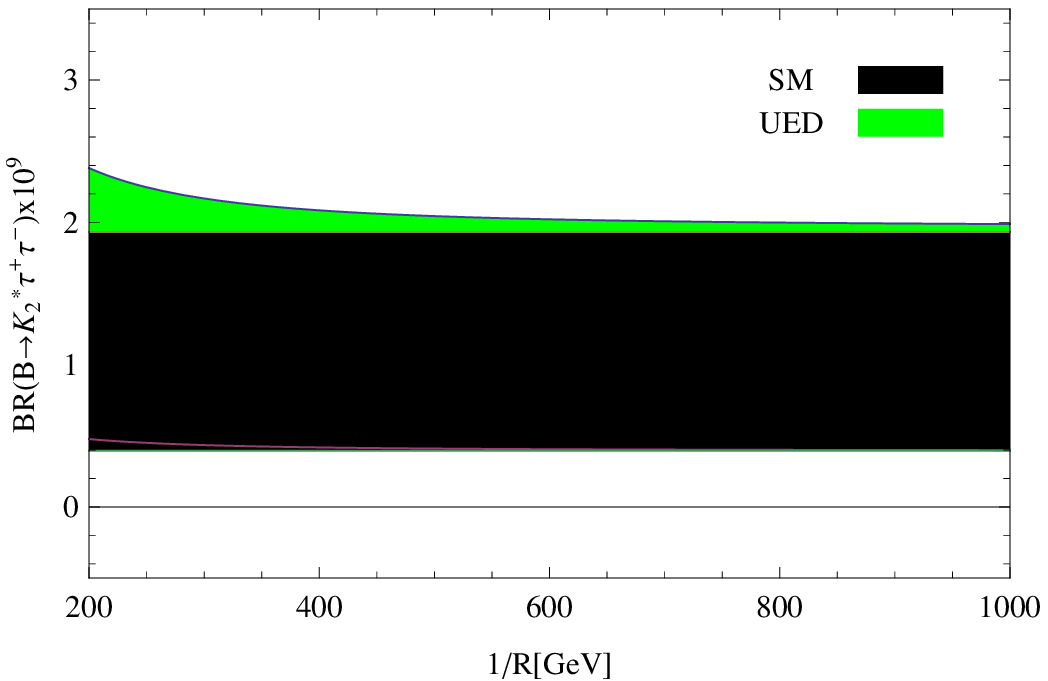,width=0.45\linewidth,clip=}
\end{tabular}
\caption{ The dependence of  branching ratios on compactification
factor, $1/R$ for $B\to K_2^*l^+l^-$}
\end{figure}
From this figure and the analysis of the branching ratios, we observe that
\begin{itemize}
\item at lower values of the compactification factor, the bands of UED obtained considering the uncertainties of  the form factors are wider compared with that of the SM   for both leptons. At higher values of $1/R$,
the two models sweep approximately the same area.

\item
 at lower values of the compactification factor $1/R$ and central values of the form factors, there is
a sizable difference between the ACD and SM model predictions for both leptons . When we increase the $1/R$, the results of UED start to diminish 
and tend to the SM predictions. The discrepancy between
the UED and SM predictions at lower values of the compactification parameter can be considered as an indication  for existence of  extra dimensions.
\item
the order of  magnitude of the branching ratio for $B\to K_2^*\mu^+\mu^-$ specially in ACD model shows  the possibility to study this channel in the  future experiments. 

\item as it is expected, an increase in the mass of final lepton results in a decrease in   the branching ratio.
\end{itemize}

\subsection{Polarization distribution}
The longitudinal polarization distribution is obtained as \cite{li}:
\begin{eqnarray}
 \frac{df_L(q^2, 1/R)}{dq^2} && = \frac{\frac{d\Gamma_0(q^2,1/R)}{dq^2}}{\frac{d\Gamma(q^2,1/R)}{dq^2}}=
\frac{3I_1^c(q^2,1/R)-I_2^c(q^2,1/R)}{3I_1^c(q^2,1/R)+6I_1^s(q^2,1/R)-I_2^c(q^2,1/R)-2I_2^s(q^2,1/R)},\nnb \\
\end{eqnarray}
where in deriving the above equation, the
\begin{eqnarray}
 \frac{ d\Gamma_0(q^2,1/R)}{dq^2}
 &=&     \bigg[|A_{L0}(q^2,1/R)|^2+|A_{R0}(q^2,1/R)|^2\bigg],
 \end{eqnarray}
 has been used for the massless limit of the differential decay width.
 The integrated polarization fraction is obtained  as \cite{li}:
\begin{eqnarray}
  {f_L(1/R)} \equiv \frac{\Gamma_0(1/R)}{\Gamma(1/R)}=\frac{\int
  dq^2\frac{d\Gamma_0(q^2,1/R)}{dq^2}}{\int
  dq^2\frac{d\Gamma(q^2,1/R)}{dq^2}}.
\end{eqnarray}
We show the sensitivity of the integrated polarization fraction to the compactification factor $1/R$ in figure 2. This figure depicts the following results:
\begin{figure}[h!]
\label{fig1} \centering
\begin{tabular}{cc}
\epsfig{file=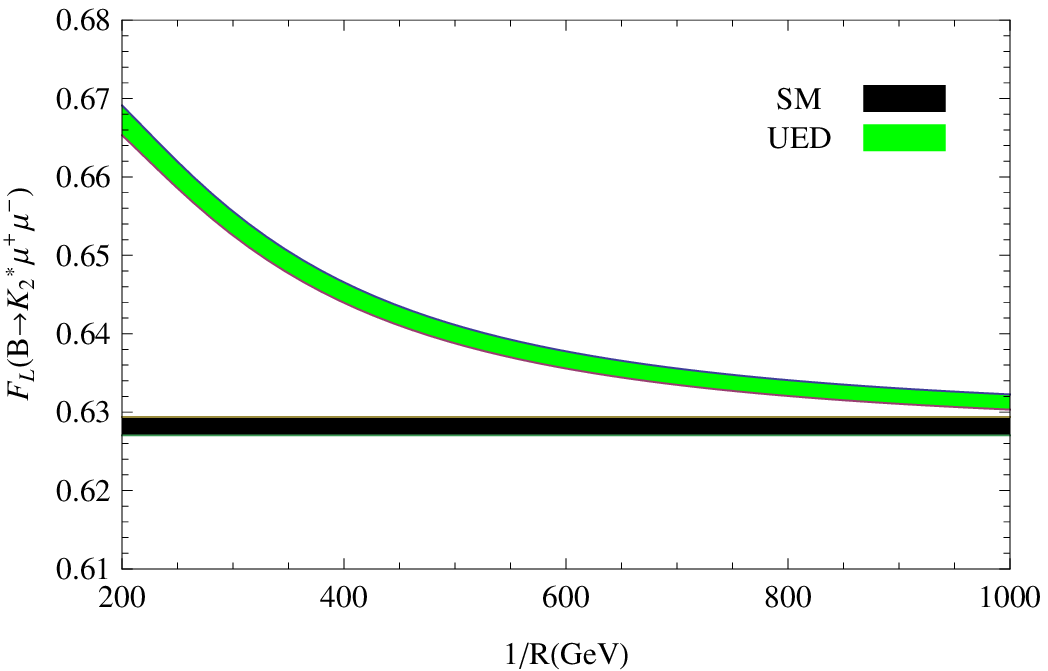,width=0.45\linewidth,clip=} &
\epsfig{file=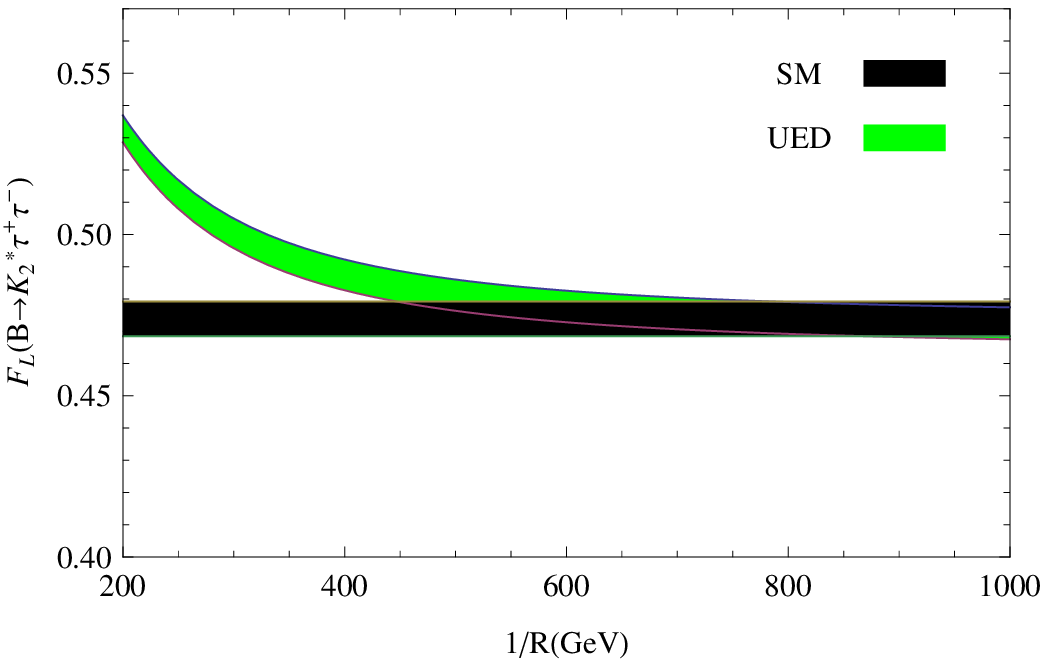,width=0.45\linewidth,clip=}
\end{tabular}
\caption{The dependence of longitudinal polarization on
compactification factor, $1/R$ for $B\to K_2^*l^+l^-$.}
\end{figure}

\begin{itemize}
\item
the UED bands deviate considerably from the SM predictions for both lepton cases at lower values of $1/R$. When the compactification factor approaches to $1000~GeV$ the
difference between the predictions of the two models  become negligible. 
\item  the errors of the form factors can not kill the discrepancies between two model predictions at lower values of the compactification factor.
\item 
when we consider the central values of the form factors, the polarization fraction for the $\mu$ case is approximately $1.4$ time greater than that of the $\tau$.
 \end{itemize}
\subsection{Forward-backward asymmetry}
The next observable related to the $B\to K_2^*l^+l^-$ transition is the forward-backward asymmetry. The differential forward-backward asymmetry  is obtained as (for details see \cite{li}):
\begin{eqnarray}
 \frac{d A_{FB}(q^2,1/R)}{dq^2}&=&\left[\int_0^1 -\int_{-1}^0\right] d\cos\theta_l\frac{d^2\Gamma(q^2,1/R)}{dq^2 d\cos\theta_l}
 =\frac{3}{4} {\cal I}(q^2,1/R),
\end{eqnarray}
where,
\begin{eqnarray}
 {\cal I}(q^2,1/R)&=& 2v
  \bigg[{\rm Re}[A_{L||}(q^2,1/R)A^*_{L\perp}(q^2,1/R)]-{\rm
  Re}[A_{R||}(q^2,1/R)A^*_{R\perp}(q^2,1/R)]\bigg].\nnb \\
\end{eqnarray}
\begin{figure}[h!]
\label{figbbbb} \centering
\begin{tabular}{cc}
\epsfig{file=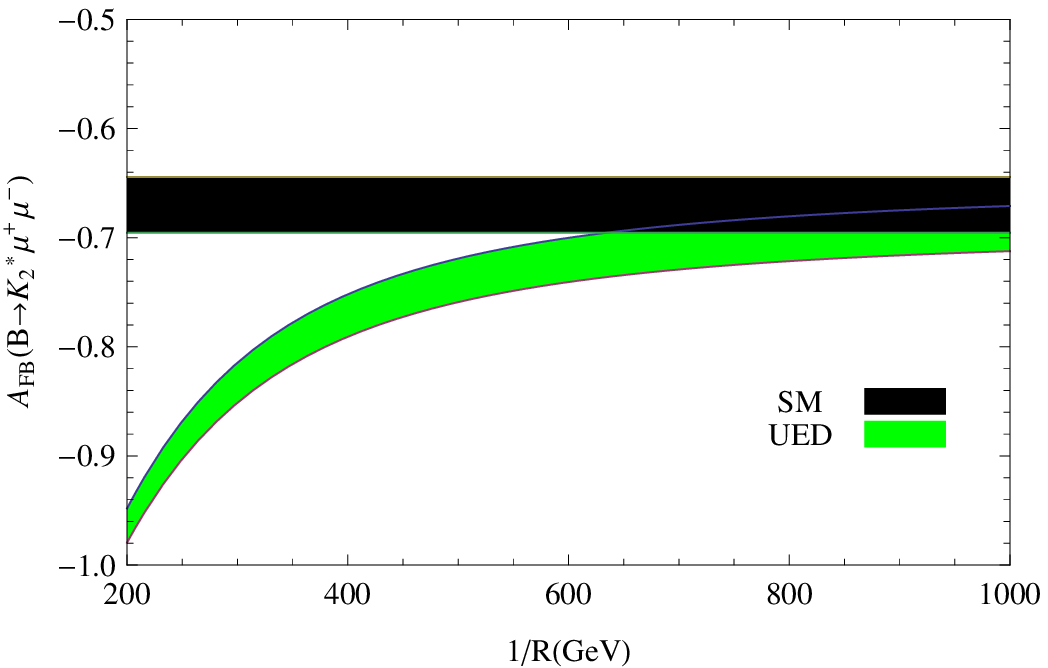,width=0.45\linewidth,clip=} &
\epsfig{file=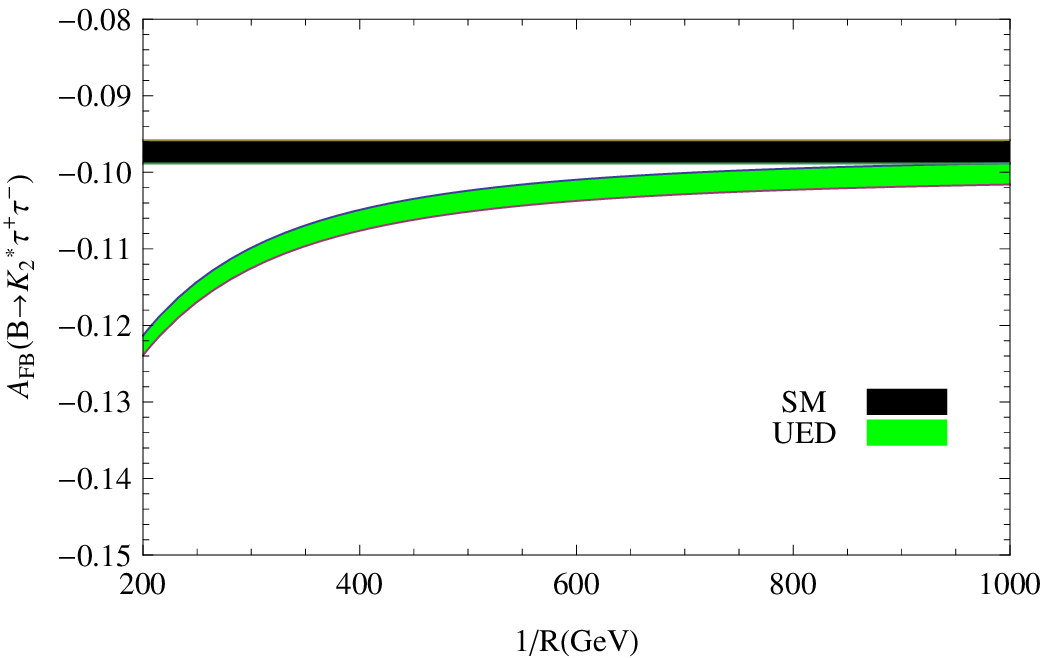,width=0.45\linewidth,clip=}
\end{tabular}
\caption{The dependence of forward-backward asymmetry on
compactification factor, $1/R$ for $B\to K_2^*l^+l^-$.}
\end{figure}
The dependence of the forward-backward asymmetry on
compactification factor for $B\to K_2^*l^+l^-$ and two leptons are shown in figure 3. From this figure and analysis of the forward-backward asymmetry, we conclude that
\begin{itemize}
\item there is a considerable discrepancy between the ACD and SM bands also in this case at lower values of the compactification factor. 
 \item as it is expected  $| A_{FB}|\leq1$ for both leptons.
\item considering the central values of the form factors, we observe that  $A_{FB}$ in the case of  $\mu$ is approximately seven times  greater than that of the $\tau$.
\end{itemize}


\section{Conclusion}
We analyzed  the rare $B\to K_2^*l^+l^-$ transition  both in ACD and 
SM models. In particular, we presented the sensitivity of some physical observables like branching ratio,  longitudinal polarization and forward-backward asymmetry
to the compactification factor, $1/R$. The order of the branching ratio of  $B\to K_2^*\mu^+\mu^-$ shows that this channel can be studied in the near future experiments.
 The obtained results show  considerable discrepancies between the prediction of the two models on the considered physical quantities at lower values of the compactification parameter. 
This discrepancy exists and can not be killed even if
the uncertainties of the form factors are taken into account. These results together with the other evidences for deviation of the
ACD model predictions from those of the SM obtained by investigation of  many observables related to the
$B$ and $\Lambda_b$ channels
 in \cite{Buras:2002ej,R7623,R7624,wangying,R7601,fazio,sirvanli,kank1,yee,bashiry,carlucci,aliev,ahmet,ek1,ek2,ek3,ek4}, can be considered as a sign for the existence of Kaluza-Klein particles
 and extra dimensions
in the nature which should we search for   at the LHC.


\begin{thebibliography}{99}
\bibitem{belle}
  J.~T.~Wei {\it et al.}  [BELLE Collaboration],
  Phys.\ Rev.\ Lett.\  {\bf 103}, 171801 (2009)
  [arXiv:0904.0770 [hep-ex]].

  \bibitem{babar}
  B.~Aubert {\it et al.}  [BABAR Collaboration],
  Phys.\ Rev.\ Lett.\  {\bf 102}, 091803 (2009)
  [arXiv:0807.4119 [hep-ex]].



\bibitem{LHC}
   B. Adeva, {\it et al.}  [LHCb Collaboration],
  arXiv:0912.4179 [hep-ex];
  M. Patel and H. Skottowe, A Fisher discriminant selection for $B_d\to K^*\mu^+\mu^-$ at LHCb,
  LHCb-2009-009.

\bibitem{superb}
  B.~O'Leary {\it et al.} [SuperB Collaboration ],
  [arXiv:1008.1541 [hep-ex]].
\bibitem{lhcb} LHCb Collaboration, Phys. Lett. B 698, 115 (2011), arXiv:1102.0206 [hep-ex].


\bibitem{antoniadis1} I. Antoniadis, Phys. Lett. B 246, 377 (1990).
\bibitem{antoniadis2} I. Antoniadis, N. Arkani, S. Dimopoulos, G. Dvali, Phys. Lett. B 439, 257 (1998).

\bibitem{arkani} N. Arkani, S. Dimopoulos, G. Dvali, Phys. Lett. B 429, 263 (1998); Phys. Rev. D 59,
086004 (1999).




\bibitem{acdd} T. Appelquist, H. C. Cheng and B. A. Dobrescu,
  Phys. Rev.  D 64, 035002 (2001).




\bibitem{Buras:2002ej}
  A.~J.~Buras, M.~Spranger and A.~Weiler,
  Nucl.\ Phys.\ B  660, 225 (2003).
\bibitem{R7623}
  A. J. Buras, A. Poschenrieder, M. Spranger
  Nucl. Phys. B  D 678, 455 (2004).
\bibitem{R7626}
  A. Buras, M. Misiak, M. M\"{u}nz and S. Pokorski,
  Nucl. Phys. B 424, 374 (1994).
\bibitem{R7627}
  M. Misiak,
  Nucl. Phys. B  393, 23 (1993);
  Erratum ibid  B  439, 161 (1995).
\bibitem{R762777}
    B. Buras, M. M\"{u}nz,
  Phys. Rev. D  52, 186 (1995).

\bibitem{wang} W. Wang, Phys. Rev. D 83, 014008 (2011).
\bibitem{zhigang} Z. G. Wang, arXiv:1011.3200 [hep-ph].
\bibitem{R7624}
  P. Colangelo, F. De Fazio, R. Ferrandes, T. N. Pham, Phys. Rev. D7 3 (2006) 115006.
\bibitem{wangying} Yu-Ming Wang, M. Jamil Aslam, Cai-Dian Lu, Eur. Phys. J. C 59, 847 (2009).
 \bibitem{R7601} T. M. Aliev, M. Savc{\i}, Eur. Phys. J. C 50, 91 (2007).
  


\bibitem{fazio} F. De Fazio, Nucl.Phys.Proc.Suppl. 174, 185-188, (2007), arXiv:hep-ph/0610208v1
\bibitem{sirvanli} B. B. Sirvanli, K. Azizi,  Y. Ipekoglu, JHEP 1101, 069 (2011).


\bibitem{kank1}
K. Azizi, N. Kat{\i}rc{\i}, JHEP 01, 087 (2011).

\bibitem{yee} T. Appelquist, H. U. Yee, Phys. Rev. D 67, 055002 (2003).
\bibitem{bashiry} V. Bashiry, M. Bayar, K. Azizi, Phys. Rev. D 78, 035010
(2008).

\bibitem{carlucci} M.V. Carlucci, P. Colangelo, F. De Fazio, Phys. Rev. D 80,
055023 (2009).

\bibitem{aliev} T. M. Aliev, M. Savci, B. B. Sirvanli, Eur. Phys. J. C 52, 375
(2007).
\bibitem{ahmet} I. Ahmed, M. A. Paracha, M. J. Aslam, Eur.
Phys. J. C 54, 591 (2008).

\bibitem{li}
 R. H. Li, C. D. L\"{u}, and W. Wang, Phys. Rev. D 83, 034034 (2011).





\bibitem{Hatanaka:2009gb}
  H.~Hatanaka and K.~C.~Yang,
  Phys.\ Rev.\  D {\bf 79}, 114008 (2009)
  [arXiv:0903.1917 [hep-ph]].

\bibitem{Hatanaka:2009sj}
  H.~Hatanaka and K.~C.~Yang,
  Eur.\ Phys.\ J.\  C {\bf 67}, 149 (2010)
  [arXiv:0907.1496 [hep-ph]].

\bibitem{Yang:2010qd}
  K.~C.~Yang,
  arXiv:1010.2944 [hep-ph].

\bibitem{Amsler:2008zz}
  K.~Nakamura  {\it et al.}  [Particle Data Group],
  J.\ Phys.\  G {\bf 37}, 075021 (2010).
\bibitem{ek1} U. Haisch and A. Weiler, Phys. Rev. D 76, 034014 (2007).
\bibitem{ek2} R. Mohanta and A. K. Giri, Phys. Rev. D 75, 035008 (2007).
\bibitem{ek3} G. Devidze, A. Liparteliani and U. G. Meissner, Phys. Lett. B 634, 59 (2006).
\bibitem{ek4} I. I. Bigi, G. G. Devidze, A. G. Liparteliani and U. G. Meissner, Phys. Rev. D 78, 097501 (2008).


\end{thebibliography}
\end{document}